\documentclass{ptephy_v1}

\usepackage{hyperref}
\usepackage{graphics}
\usepackage{bm}

\usepackage{color} 
\usepackage{soul} 

\begin{document}

\title{Magnetic Field Induced by Straight Currents on the Hyperboloid}

\author{Roman Kr\v{c}mar$^{1,2}_{~}$}
\author{Andrej Gendiar$^{1,2}_{~}$}
\author{Tomotoshi Nishino$^{2*}_{~}$ \email{nishino@kobe-u.ac.jp} }

\affil{$^1_{~}$Institute of Physics, Slovak Academy of Sciences, 
D\'ubravska cesk\'a 9, SK-845 11, Bratislava, Slovakia}
\affil{$^2_{~}$Department of Physics, Graduate School of Science, Kobe University, 
Kobe 657-8501, Japan}

\begin{abstract}%
We consider the magnetic field induced by the steady or the quasi-steady electric currents 
that flow along the straight wires, which are equidistantly arranged on the hyperboloid. 
The spatial distribution of the magnetic field and the force acting on each wire are
calculated. The continuum limit, where the wires are densely aligned, is also considered. 
We discuss the application of the hyperbolic current configuration to the generation 
of high magnetic fields.
\end{abstract}

\subjectindex{xxxx, xxx}

\maketitle

\section{Introduction}

Walking along the seaside of Kobe city in Japan, one encounters the famous red tower, 
which is called {\it The Kobe Port Tower}, shown in Fig.~1. 
It is 108 m high, and was constructed in 1963.
The architectural structure is formed by 32 straight, jointed pipes, 
which typically exhibit the hyperbolic envelope surface. 
In this article, we consider the generation of a magnetic field by means of
the hyperbolic structure. 

%
\begin{figure}[h]
\centering\includegraphics[width = 8.0 cm]{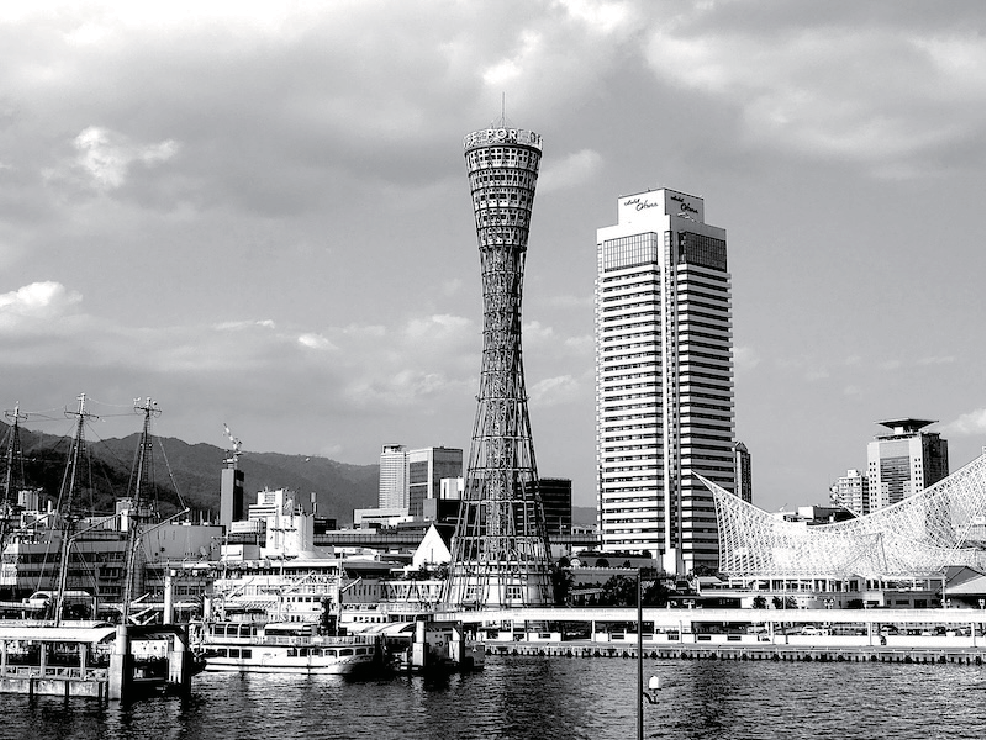}
\caption{{\it The Kobe Port Tower}, which is well known by the hyperbolic structure.}
\end{figure}

When the uniform electric current of the surface density $\sigma$ flows on 
the surface of the infinitely long cylinder of radius $R$ in parallel with the central axis, 
the Maxwell's equation~\cite{Maxwell} tells that the surrounding magnetic field 
exists outside the cylinder $r > R$, where $r$ is the distance from 
the axis. The corresponding field strength is represented as
\begin{equation}
H( r ) = \sigma \frac{R}{r}  \, 
\end{equation}
for $r > R$, and there is no field inside $r < R$. 
As a result, the electromagnetic force per area $f = \mu \sigma^2_{~} / 2$ acts on the 
surface current, where $\mu$ represents the permeability of the medium, to the 
direction of shrinking the cylinder. The force strength $f$ can be 
extracted from the discontinuity of the Maxwell stress at $r = R$.

\begin{figure}[h]
\centering\includegraphics[width = 4.7 cm]{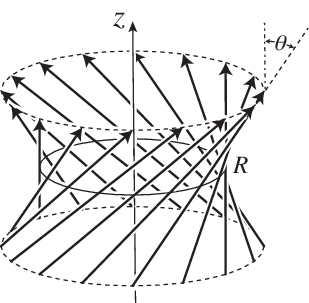}
\caption{Straight current flow on the hyperboloid, which is specified by Eq.~(2). 
Only the finite area 
near the $xy$ plane is drawn from the infinitely extending lines.}
\end{figure}

Let us consider the one-parameter deformation applied to the cylinder. 
Suppose that the current flows in a straight manner on the hyperboloid, 
as pictorially shown in Fig.~2. The curved surface is specified by the relation
\begin{equation}
x^2_{~} + y^2_{~} - \left( z \tan\theta \right)^2_{~} = R^2_{~} \, ,
\end{equation}
where $x$, $y$, and $z$ represent the Cartesian coordinates. 
The deformation parameter $\theta$ represents the tilted angle $\theta$ of the lines 
with respect to the $z$-axis.~\cite{cosh} 
When $\theta = 0$, the hyperboloid reduces to the cylinder of radius $R$. 
Throughout this article, we assume that the hyperboloid extends to infinity ($-\infty < z < \infty$). 
In the following, we assume that the current flows along the straight wires 
that are equidistantly arranged on the hyperboloid, and calculate the magnetic 
filed inside the hyperboloid $x^2_{~} + y^2_{~} - \left( z \tan\theta \right)^2_{~} < R^2_{~}$, 
and the outside $x^2_{~} + y^2_{~} - \left( z \tan\theta \right)^2_{~} > R^2_{~}$. 
We observe the compensation of the electromagnetic force acting on the wires, 
in particular when $\theta = \pi / 4$. We then consider the continuum limit. 

In the next section, we represent the quasi-uniform current on the hyperboloid by 
the assembly of discrete currents, which flow through the straight wires. The corresponding 
magnetic field is calculated, in particular along the central axis of the hyperboloid. 
In Sec.~3, we observe the electromagnetic force acting on each wire. 
In the last section, we conclude the obtained results and discuss a possible application 
of the considered geometrical setup as a high-field magnet.

\section{Straight Currents on a Hyperboloid}

We first set up the discrete current flow on the hyperboloid, which is specified by Eq.~(2). 
Suppose that $N$ numbers of straight conducting wires are arranged equidistantly on 
the hyperboloid, as shown in Fig.~2, where the case $N = 16$ is typically 
drawn. Let us label the wires by the index $\ell = 0, \, 1, \, 2, \, \cdots$, and $N-1$.
On the left side of Fig.~3, we see the location of the $\ell$-th wire, which is tilted by the angle 
$\theta$ from the $z$-axis toward the direction of the azimuth angle 
$\phi_{\ell}^{~} = 2\pi \ell / N$ with respect to the $x$-axis.  
The crossing point P$\!_{\ell}^{~}$ of the wire with respect to the $xy$-plane is expressed 
by the coordinates $x = R \sin\phi_{\ell}^{~}$, $y = - R \cos\phi_{\ell}^{~}$, and $z = 0$. 

\begin{figure}[h]
\centering\includegraphics[width = 11.0 cm]{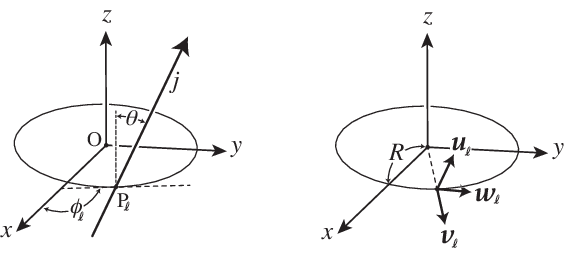}
\caption{
Left: Location of the $\ell$-th wire, which passes through the point P$\!_{\ell}^{~}$ on 
the $xy$-plane. Right: Orthonormal vectors ${\bm u}_{\ell}^{~}$, ${\bm v}_{\ell}^{~}$, 
and ${\bm w}_{\ell}^{~}$ defined in Eq.~(4).}
\end{figure}

Any point on the $\ell$-th wire can be represented by the position vector
\begin{equation}
{\bm q}_{\ell}^{~}[ \alpha ] = 
\left( \begin{matrix}
 \alpha \sin\theta \cos\phi_{\ell}^{~} + R \sin\phi_{\ell}^{~}  \\
 \alpha \sin\theta \sin\phi_{\ell}^{~} - R \cos\phi_{\ell}^{~} \\
 \alpha \cos\theta ~~~~~~~~~~~~~~~~~~~~~ \,
\end{matrix} \right) \, ,
\end{equation}
where the parameter $\alpha$ can be interpreted as the one-dimensional coordinate on 
the wire. It is straightforward to check that the coordinates satisfy the relation in Eq.~(2). 
For the latter convenience, we introduce the set of orthonormal vectors
\begin{equation}
{\bm u}_{\ell}^{~} = 
\left( \begin{matrix}
 \sin\theta \cos\phi_{\ell}^{~} \\  \sin\theta \sin\phi_{\ell}^{~} \\ \cos\theta ~~~~~~~ \,
\end{matrix} \right) \, , ~~~
{\bm v}_{\ell}^{~} = 
\left( \begin{matrix} ~~ \sin\phi_{\ell}^{~} \\ - \cos\phi_{\ell}^{~} \\ ~ 0
\end{matrix} \right) \, , ~~~
{\bm w}_{\ell}^{~} = {\bm u}_{\ell}^{~} \times {\bm v}_{\ell}^{~} = 
\left( \begin{matrix}
~~ \cos\theta \cos\phi_{\ell}^{~}  \\
~~ \cos\theta \sin\phi_{\ell}^{~}  \\
- \sin\theta ~~~~~~~~
\end{matrix} \right) \, ,
\end{equation}
which enables us to express ${\bm q}_{\ell}^{~}[ \alpha ]$ in the linear form
\begin{equation}
{\bm q}_{\ell}^{~}[ \alpha ] = \alpha {\bm u}_{\ell}^{~} + R {\bm v}_{\ell}^{~} \, .
\end{equation}
%
We assume that the current of the magnitude $j$ flows through each wire in the direction of 
${\bm u}_{\ell}^{~}$. The corresponding current vector is represented by 
\begin{equation}
{\bm j}_{\ell}^{~} = j {\bm u}_{\ell}^{~} \, . 
\end{equation}
\begin{figure}[h]
\centering\includegraphics[width = 2.6 cm]{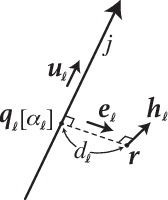}
\caption{The distance $d_{\ell}^{~}$ from the location ${\bm r}$ to the $\ell$-th wire.}
\end{figure}

Let us calculate the magnetic field at an arbitrary position ${\bm r}$, 
whose components are $x$, $y$, and $z$, induced by the current ${\bm j}_{\ell}^{~}$. 
We assume that the wire is infinitely long.
In order to obtain the distance $d_{\ell}^{~}$ from the point ${\bm r}$ to the $\ell$-th wire, 
we first determine the root of the perpendicular line from ${\bm r}$ to the wire. 
The position of the root ${\bm q}_{\ell}^{~}\left[ {\alpha}_{\ell}^{~} \right]$ 
shown in Fig.~4 satisfies the orthogonal relation
\begin{equation}
\left( {\bm r} - {\bm q}_{\ell}^{~}[ {\alpha}_{\ell}^{~} ] \right) \cdot {\bm u}_{\ell}^{~} = 
\left( {\bm r}  - {\alpha}_{\ell}^{~}{\bm u}_{\ell}^{~} - R{\bm v}_{\ell}^{~} \right) 
\cdot {\bm u}_{\ell}^{~} = 0 \, ,
\end{equation}
which determines the specific parameter
\begin{equation}
{\alpha}_{\ell}^{~} = {\bm r} \cdot {\bm u}_{\ell}^{~} =
x \sin\theta \cos\phi_{\ell}^{~} + 
y \sin\theta \sin\phi_{\ell}^{~} + 
z \cos\theta \, .
\end{equation}
For convenience, we also define the following parameters
\begin{align}
{\beta}_{\ell}^{~} &= {\bm r} \cdot {\bm v}_{\ell}^{~} =
x \sin\phi_{\ell}^{~} -
y \cos\phi_{\ell}^{~} \, ,\nonumber\\
{\gamma}_{\ell}^{~} &= {\bm r} \cdot {\bm w}_{\ell}^{~} =
 x \cos\theta \cos\phi_{\ell}^{~} 
+ y \cos\theta \sin\phi_{\ell}^{~} 
- z \sin\theta \, ,
\end{align}
and represent ${\bm r}$ in the form
$
{\bm r} = 
{\alpha}_{\ell}^{~} {\bm u}_{\ell}^{~} + 
{\beta}_{\ell}^{~} {\bm v}_{\ell}^{~} + 
{\gamma}_{\ell}^{~} {\bm w}_{\ell}^{~}
$.
We then have the relation
\begin{equation}
{\bm r} - {\bm q}_{\ell}^{~}[ {\alpha}_{\ell}^{~} ] =
\left( {\beta}_{\ell}^{~} - R \right){\bm v}_{\ell}^{~} + {\gamma}_{\ell}^{~} {\bm w}_{\ell}^{~} = 
d_{\ell}^{~} {\bf e}_{\ell}^{~} \, ,
\end{equation}
where the distance $d_{\ell}^{~}$ is expressed as
\begin{equation}
d_{\ell}^{~} = \sqrt{ \left( {\beta}_{\ell}^{~} - R \right)^2_{~} + {\gamma}_{\ell}^{\, 2} }  \, ,
\end{equation}
and the unit vector ${\bm e}_{\ell}^{~}$ is defined as
\begin{equation}
{\bm e}_{\ell}^{~} = \frac{{\beta}_{\ell}^{~} - R}{d_{\ell}^{~}} \, {\bm v}_{\ell}^{~} + 
\frac{{\gamma}_{\ell}^{~}}{d_{\ell}^{~}} \, {\bm w}_{\ell}^{~} \, .
\end{equation}
As a consequence, the magnetic field created by ${\bm j}_{\ell}^{~}$ at the position ${\bm r}$ is
obtained as
\begin{equation}
{\bm h}_{\ell}^{~}( {\bm r} ) = \frac{j}{2 \pi d_{\ell}^{~}} \, {\bm u}_{\ell}^{~} \times {\bm e}_{\ell}^{~} =
j  \frac{{\beta}_{\ell}^{~} - R}{2 \pi d_{\ell}^{\, 2}} \, {\bm w}_{\ell}^{~} -
j \frac{{\gamma}_{\ell}^{~}}{2 \pi d_{\ell}^{\, 2}} \, {\bm v}_{\ell}^{~} \, .
\end{equation}
It should be noted that we ignore the thickness of the wires throughout this article. 
Summing up the contributions from all the wires, we obtain the magnetic field 
\begin{equation}
{\bm H}( {\bm r} ) = \sum_{\ell = 0}^{N-1} {\bm h}_{\ell}^{~}( {\bm r} ) =
\frac{j}{2\pi} \sum_{\ell = 0}^{N-1} 
\frac{{\beta}_{\ell}^{~} - R}{\left( {\beta}_{\ell}^{~} - R \right)^2_{~} + 
{\gamma}_{\ell}^{\, 2} } \, {\bm w}_{\ell}^{~} -
\frac{j}{2\pi} \sum_{\ell = 0}^{N-1} 
\frac{{\gamma}_{\ell}^{~}}{\left( {\beta}_{\ell}^{~} - R \right)^2_{~} + 
{\gamma}_{\ell}^{\, 2}  } \, {\bm v}_{\ell}^{~}
\end{equation}
at the position ${\bm r}$. 

Let us check the field along the $z$-axis. Since the conditions ${\beta}_{\ell}^{~} = 0$ and 
${\gamma}_{\ell}^{~} = - z \sin\theta$ are satisfied on the axis, we obtain a simpler form
\begin{equation}
{\bm H}( z {\bm e}_z^{~} ) = 
\sum_{\ell = 0}^{N-1} {\bm h}_{\ell}^{~}( z {\bm e}_z^{~} ) =
\frac{j}{2\pi} \sum_{\ell = 0}^{N-1} 
\frac{ - R}{R^2_{~} + \left( z \sin\theta \right)^{2}_{~} } \, {\bm w}_{\ell}^{~} -
\frac{j}{2\pi} \sum_{\ell = 0}^{N-1} 
\frac{- z \sin\theta}{R^2_{~} + \left( z \sin\theta \right)^{2}_{~}  } \, {\bm v}_{\ell}^{~} \, ,
\end{equation}
where ${\bm e}_z^{~}$ is the unit vector that points the direction of the $z$-axis. 
Remembering that $\theta_{\ell}^{~} = 2 \pi \ell / N$, the relations
\begin{equation}
\sum_{\ell = 0}^{N-1} {\bm v}_{\ell}^{~} = {\bm 0} \, , ~~~~~~~~~
\sum_{\ell = 0}^{N-1} {\bm w}_{\ell}^{~} = - N \sin\theta \, {\bm e}_z^{~} 
\end{equation}
holds in Eq.~(15), and thus we further obtain
\begin{equation}
{\bm H}( z {\bm e}_z^{~} ) = 
\frac{N j}{2\pi} \, \frac{ R \sin\theta}{ R^2_{~} + \left( z \sin\theta  \right)^{2}_{~} } \, {\bm e}_z^{~} = 
\frac{N j}{2\pi R} \, \frac{ \sin\theta}{ 1 + \bigl[ \, ( z / R ) \sin\theta \, \bigr]^{2}_{~} } \, {\bm e}_z^{~} \, .
\end{equation}
It should be noted that ${\bm H}( z {\bm e}_z^{~} )$ is written by the analytic function, 
and the $N$-dependence only appears as the pre-factor.

The magnitude of the field on the $z$-axis $H_{\rm c}^{~}( z ) = 
\bigl| {\bm H}( z {\bm e}_z^{~} ) \bigr|$ is strongest at the origin $z = 0$,
\begin{equation}
H_{\rm c}^{~}( 0 ) = \frac{ N j \sin\theta  }{2\pi R} \, ,
\end{equation}
and decays asymptotically with $z^{-2}_{~}$ in the region $| z | \gg R$. 
The value $N j \sin\theta$ corresponds to the averaged horizontal current density on the 
hyperboloid at $z = 0$, on the cross section with the $xy$-plane.  

\begin{figure}[h]
\centering\includegraphics[width = 8.5 cm]{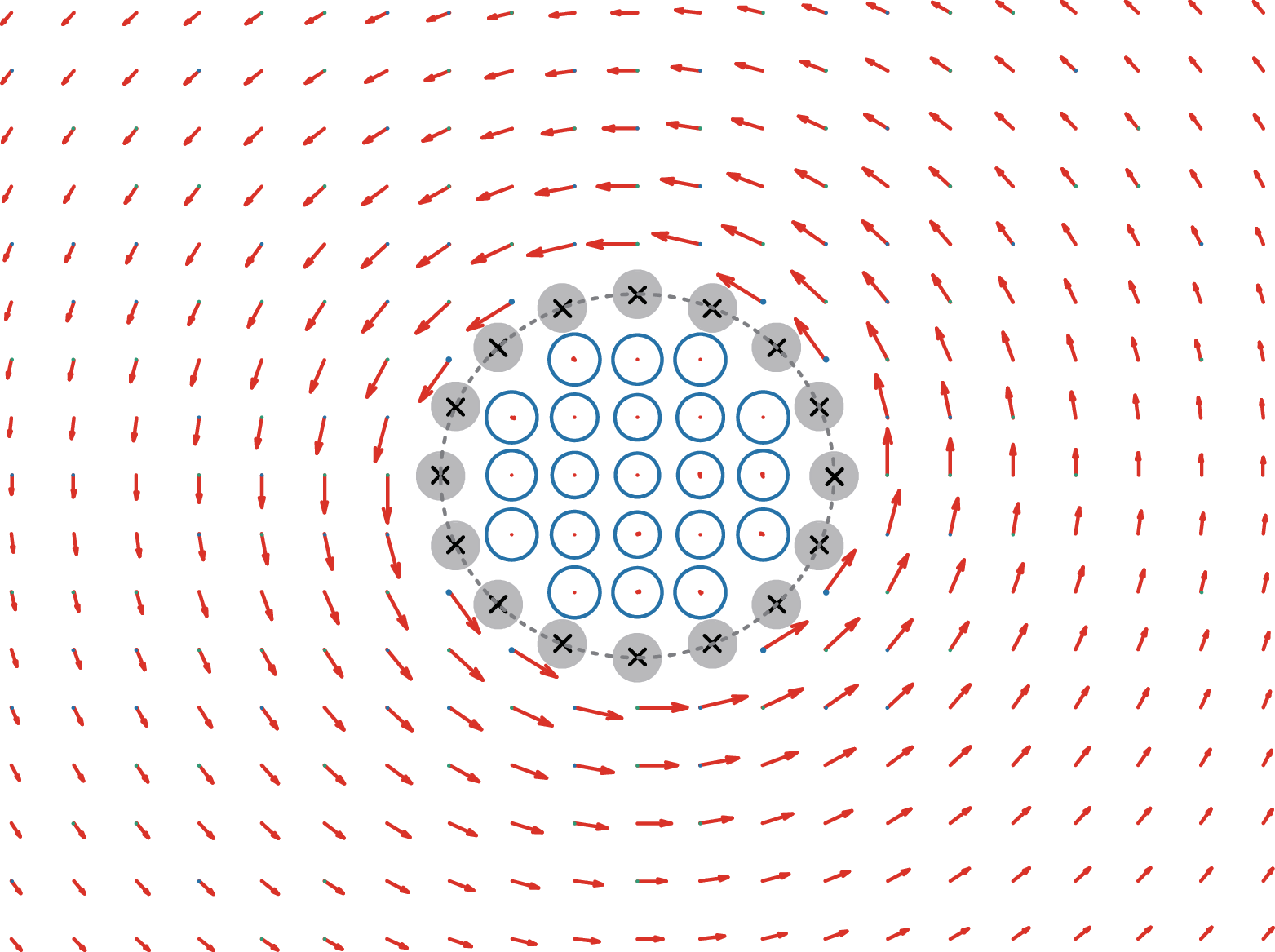}
\caption{Field configuration on the $xy$-plane, when $N = 16$ and $\theta = \pi / 4$. 
Components parallel to the $xy$-plane are shown by the red arrows, 
and the $z$-components are shown by blue circles. Inside the gray area around
the crossing points of the wires, which are shown by cross marks, the arrows and
circles are not drawn. }
\end{figure}

Let us observe the magnetic field under the condition $N = 16$ and 
$\theta = \pi / 4$, for the purpose of capturing the spatial arrangement of the field.
We consider the typical strength $H_{\rm c}^{~}( 0 )$ as the unit of magnetic field; therefore, $j$ and $R$ can be chosen arbitrarily, provided that they are positive.
Figure 5 shows the field configuration on the $xy$-plane, where $z = 0$. 
The red arrows indicate the direction of the $xy$-component at the lattice points shown 
by the dots, and the arrow length is proportional to the horizontal magnitude 
${\displaystyle \sqrt{ 
\bigl[ H_x^{~}( {\bm r} ) \bigr]^2_{~} + 
\bigl[ H_y^{~}( {\bm r} ) \bigr]^2_{~} } }$.
The $z$ components $H_z^{~}( {\bm r} )$ are shown by the blue circles, 
where the size of the marks are proportional to ${\displaystyle \bigl| H_z^{~}( {\bm r} ) \bigr|}$. 
We have chosen the mark size so that they do not overlap conspicuously; note that the blue circle at the center denotes $H_{\rm c}^{~}( 0 )$. 
To avoid the complexity, we do not draw arrows and circles inside the gray area, 
which is the vicinity of the crossing point of the wires with respect to the $xy$-plane, 
drawn by cross marks. The area can be considered as the place where 
metal wires of finite thickness pass through under a realistic experimental setup. 
Inside the hyperboloid, the field is almost parallel to the $z$-axis and nearly uniform, 
where the strength is approximately $H_{\rm c}^{~}( 0 )$ in Eq.~(18). Outside the hyperboloid, 
there is a horizontally surrounding field, whose strength can be roughly estimated as 
${\displaystyle N j \cos\theta / \left| {\bm r} \right| }$, where $N j \cos\theta$ is the 
$z$-component of the total current. Scattered behavior is limited in the neighborhood 
of the wires, typically up to the distance between nearest wires. 

\begin{figure}[h]
\centering\includegraphics[width = 9.5 cm]{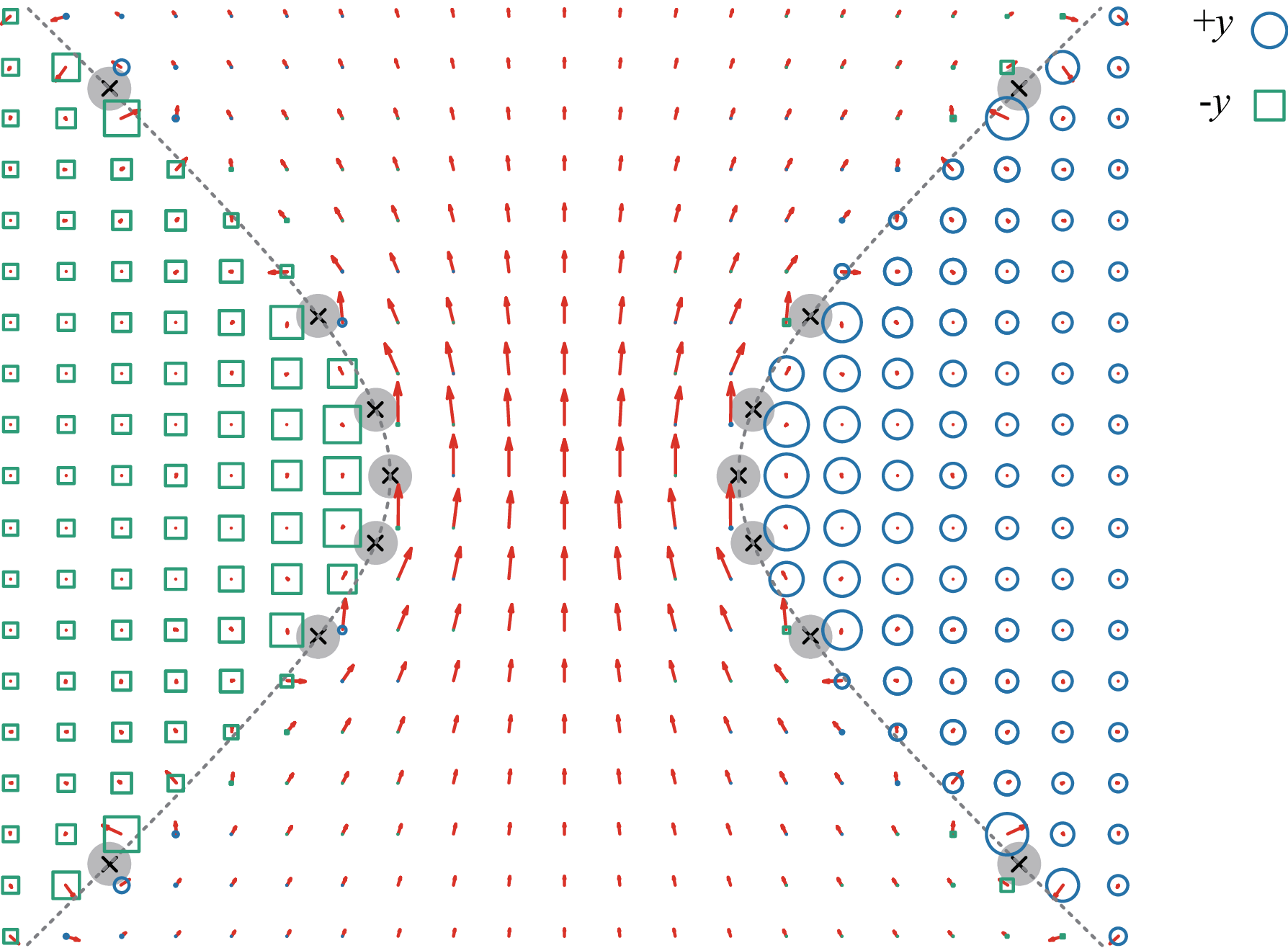}
\caption{Field configuration on the $xz$-plane, when $N = 16$ and $\theta = \pi / 4$. 
Components parallel to the $xz$-plane are shown by arrows, and
the $y$-components are shown by circles when they are positive, 
and squares when negative. Inside the gray area around
the crossing points of the wires shown by cross marks, 
the field configurations are not drawn.}
\end{figure}

Figure 6 shows the field configurations on the $xz$-plane, where $y = 0$, under the same 
conditions as in Fig.~5. The dotted curves represent the cross-section of the hyperboloid. 
The $xz$-components are drawn by red arrows, and the $y$ components are 
indicated by blue circles when they are positive, and by green squares when they are 
negative. In the same manner as Fig.~5, we skip drawing marks in the vicinity of the 
crossing point of the wire shown by the gray area. We see that the field is strong 
around {\it the neck} of the hyperboloid, and continuously decays with increasing 
$\left| {\bm r} \right|$. On the $z$-axis, the asymptotic decay is proportional to $z^{-2}_{~}$, 
and to $| x |^{-1}_{~}$ on the $x$-axis. 

When $N$ is sufficiently large, we can approximately express ${\bm H}( {\bm r} )$ by the 
continuum integral forms. For example, we have on the $xz$-plane
\begin{align}
H_{x}^{~}( {\bm r} ) &= 
\frac{N j}{( 2\pi )^2_{~}} \int_0^{2\pi} \frac{- R \cos\theta \cos\phi + z \sin\theta \sin\phi
}{\left( x \sin \phi - R \right)^2_{~} + \left( x \cos\theta \cos\phi - z \sin\theta \right)^2_{~} } \,\, d\phi 
\, , \\
H_{y}^{~}( {\bm r} ) &= 
\frac{N j}{( 2\pi )^2_{~}} \int_0^{2\pi} 
\frac{ x \cos\theta - R \cos\theta \sin\phi - z \sin\theta \cos\phi
}{\left( x \sin \phi - R \right)^2_{~} + \left( x \cos\theta \cos\phi - z \sin\theta \right)^2_{~} } \,\, d\phi 
\, , \\
H_{z}^{~}( {\bm r} ) &= 
\frac{N j}{( 2\pi )^2_{~}} \int_0^{2\pi} \frac{- x \sin\theta\sin\phi + R \sin\theta 
}{\left( x \sin \phi - R \right)^2_{~} + \left( x \cos\theta \cos\phi - z \sin\theta \right)^2_{~} } \,\, d\phi 
\, , 
\end{align} 
where we have substituted the simpler relations $\beta = x \sin \phi$ 
and $\gamma = x \cos\theta \cos\phi - z \sin\theta$. In the continuum limit, the rotational 
symmetry with respect to the $z$-axis tells that the field outside the hyperboloid on the 
$xy$-plane can be expressed as
\begin{equation}
{\bm H}( {\bf r} ) = - \frac{Nj \cos\theta}{2 \pi r}  \, {\bf e}_z^{~} \times {\bf e}_r^{~} 
= - \frac{Nj \cos\theta}{2 \pi r}  \, {\bf e}_{\theta}^{~}  \, ,
\end{equation}
where $r = | {\bm r} | = \sqrt{ x^2_{~} + y^2_{~} }$, ${\bm e}_r^{~} = {\bm r} / r$, 
and ${\bf e}_{\phi}^{~} = {\bf e}_z^{~} \times {\bf e}_r^{~} $. 
It is easy to check that the field is actually independent of $z$ outside the hyperboloid.
Inside the hyperboloid and on the $xy$-plane, the field strength is equal to 
$H_{\rm c}^{~}( 0 ) = Nj \sin\theta / 2 \pi R$. Thus in the specific case $\theta = \pi / 4$, 
the $H_{\rm c}^{~}( 0 )$ is equal to the field strength just outside $Nj \cos\theta / 2 \pi R$, 
where $r = R_{+}^{~}$. This means that the Maxwell stress is balanced on the crossing 
section of the hyperboloid with respect to the $xy$-plane.

\section{Force distribution}

Let us define the magnetic field created by the wires other than the $0$-th one
\begin{equation}
{\overline {\bm H} }( {\bf r} ) = \sum_{\ell = 1}^{N-1} {\bm h}_{\ell}^{~}( {\bm r} ) \, .
\end{equation}
The electromagnetic force per length acting on the $0$-th wire 
at the position ${\bm q}_{0}^{~}[ \alpha ]$ is then expressed as
\begin{align}
{\bm f}_0^{~}[ \alpha ] &= 
\mu {\bm j}_{0}^{~}  \times {\overline {\bm H} }\left( {\bm q}_0^{~}[ \alpha ] \right) = 
\mu j \sum_{\ell = 1}^{N-1} {\bm u}_{0}^{~} \times 
{\bm h}_{\ell}^{~}\left( {\bm q}_0^{~}[ \alpha ] \right) \nonumber\\
&= 
\frac{\mu j^2_{~}}{2\pi} \sum_{\ell = 1}^{N-1} 
\frac{{\beta}_{\ell}^{~} - R}{\left( {\beta}_{\ell}^{~} - R \right)^2_{~} + 
{\gamma}_{\ell}^{\, 2} } \, {\bm u}_{0}^{~} \times {\bm w}_{\ell}^{~} -
\frac{\mu j^2_{~}}{2\pi} \sum_{\ell = 1}^{N-1} 
\frac{{\gamma}_{\ell}^{~}}{\left( {\beta}_{\ell}^{~} - R \right)^2_{~} + 
{\gamma}_{\ell}^{\, 2}  } \, {\bm u}_{0}^{~} \times {\bm v}_{\ell}^{~} \, .
\end{align}
%
A simple case is $\alpha = 0$, where the components of ${\bm q}_0^{~}[ 0 ]$ 
are $x = 0$, $y = - R$, and $z = 0$. The corresponding parameters are
$\beta_{\ell}^{~} = R \cos\phi_{\ell}^{~}$ and 
$\gamma_{\ell}^{~} = - R \cos\theta\sin\phi_{\ell}^{~}$, and the denominator
\begin{equation}
d_{\ell}^2 = 
\left( {\beta}_{\ell}^{~} - R \right)^2_{~} + {\gamma}_{\ell}^{\, 2} = 
R^2_{~} \left( \cos\phi_{\ell}^{~} - 1 \right)^2_{~} + 
R^2_{~} \left( \cos\theta\sin\phi_{\ell}^{~} \right)^2_{~}
\end{equation}
is invariant under the sign change $\phi_{\ell}^{~} \rightarrow - \phi_{\ell}^{~}$. 
Looking at the components of the vectors
\begin{align}
{\bm u}_{0}^{~} \times {\bm w}_{\ell}^{~} &= 
\left( \begin{matrix} \sin\theta \\ 0 \\ \cos\theta \end{matrix} \right) \times 
\left( \begin{matrix} 
~~~ \cos\theta\cos\phi_{\ell}^{~} \\ 
~~~ \cos\theta\sin\phi_{\ell}^{~} \\
-\sin\theta ~~~~~~~ \end{matrix} \right) 
= 
\left( \begin{matrix} 
~~~~~~~~~~~ \,  - (\cos\theta)^2_{~}\sin\phi_{\ell}^{~} \\ 
(\sin\theta)^2_{~} + (\cos\theta)^2_{~} \cos\phi_{\ell}^{~} \\ 
~~~~~~~~~~~ \sin\theta\cos\theta\sin\phi_{\ell}^{~} 
\end{matrix} \right) \, , \\
{\bm u}_{0}^{~} \times {\bm v}_{\ell}^{~} &=
\left( \begin{matrix} \sin\theta \\ 0 \\ \cos\theta \end{matrix} \right) \times
\left( \begin{matrix} ~~~ \sin\phi_{\ell}^{~} \\ - \cos\phi_{\ell}^{~} \\ 0 \end{matrix} \right) 
= 
\left( \begin{matrix} 
~~ \cos\theta\cos\phi_{\ell}^{~} \\ 
~~ \, \cos\theta\sin\phi_{\ell}^{~} \\ 
-\sin\theta\cos\phi_{\ell}^{~} 
\end{matrix} \right) \, ,
\end{align}
and considering the parity with respect to $\phi_{\ell}^{~} = 2 \pi \ell / N$, 
one finds that the $x$- and the $z$-components of the force vanish. 
The fact can be checked from the discrete rotational symmetry of the system.
The $y$-component of the force
\begin{equation}
\frac{\mu j^2_{~}}{2\pi R} \sum_{\ell = 1}^{N-1} 
\frac{ \left( \cos\phi_{\ell}^{~} - 1 \right) \left[ (\sin\theta)^2_{~} + (\cos\theta)^2_{~} \cos\phi_{\ell}^{~} \right] +
\left( \cos\theta\sin\phi_{\ell}^{~} \right)^2_{~}
}{ \left( \cos\phi_{\ell}^{~} - 1 \right)^2_{~} + \left( \cos\theta\sin\phi_{\ell}^{~} \right)^2_{~} } 
\end{equation}
is dependent on $\theta$. It should be noted that when $\theta = \pi / 4$, 
the cancellation 
\begin{equation}
\frac{\mu j^2_{~}}{2\pi R} \sum_{\ell = 1}^{N-1} 
\frac{(1/2) \left( \cos\phi_{\ell}^{~} - 1 \right) \left( 1 + \cos\phi_{\ell}^{~} \right) +
(1/2) \left( \sin\phi_{\ell}^{~} \right)^2_{~}
}{\left( \cos\phi_{\ell}^{~} + 1 \right)^2_{~} + (1/2) \left( \sin\phi_{\ell}^{~} \right)^2_{~} } = 0 
\end{equation}
happens in each term, and the $y$-component also vanishes. Thus, in this case, no force act to the wires at the crossing point with the $xy$-plane.

When $\theta = \pi / 4$, the force ${\bm f}_0^{~}[ \alpha ]$ is actually kept weak even 
for $\alpha \neq 0$. Figure 7 shows the components 
${\bm f}_0^{~}[ \alpha ] \cdot {\bm v}_0^{~}$ and 
${\bm f}_0^{~}[ \alpha ] \cdot {\bm w}_0^{~}$, respectively, by circles and squares, 
in the unit $f^{*}_{~} = \mu N j^2_{~} / 2 \pi R$, with respect to $\alpha$, regarding $R$ 
as the unit of length. The residual force of the order of $0.2$ around 
$\alpha / R \sim 2$ exists even for larger $N$, and is not the finite $N$ effect.

\begin{figure}[h]
\centering\includegraphics[width = 9.0 cm]{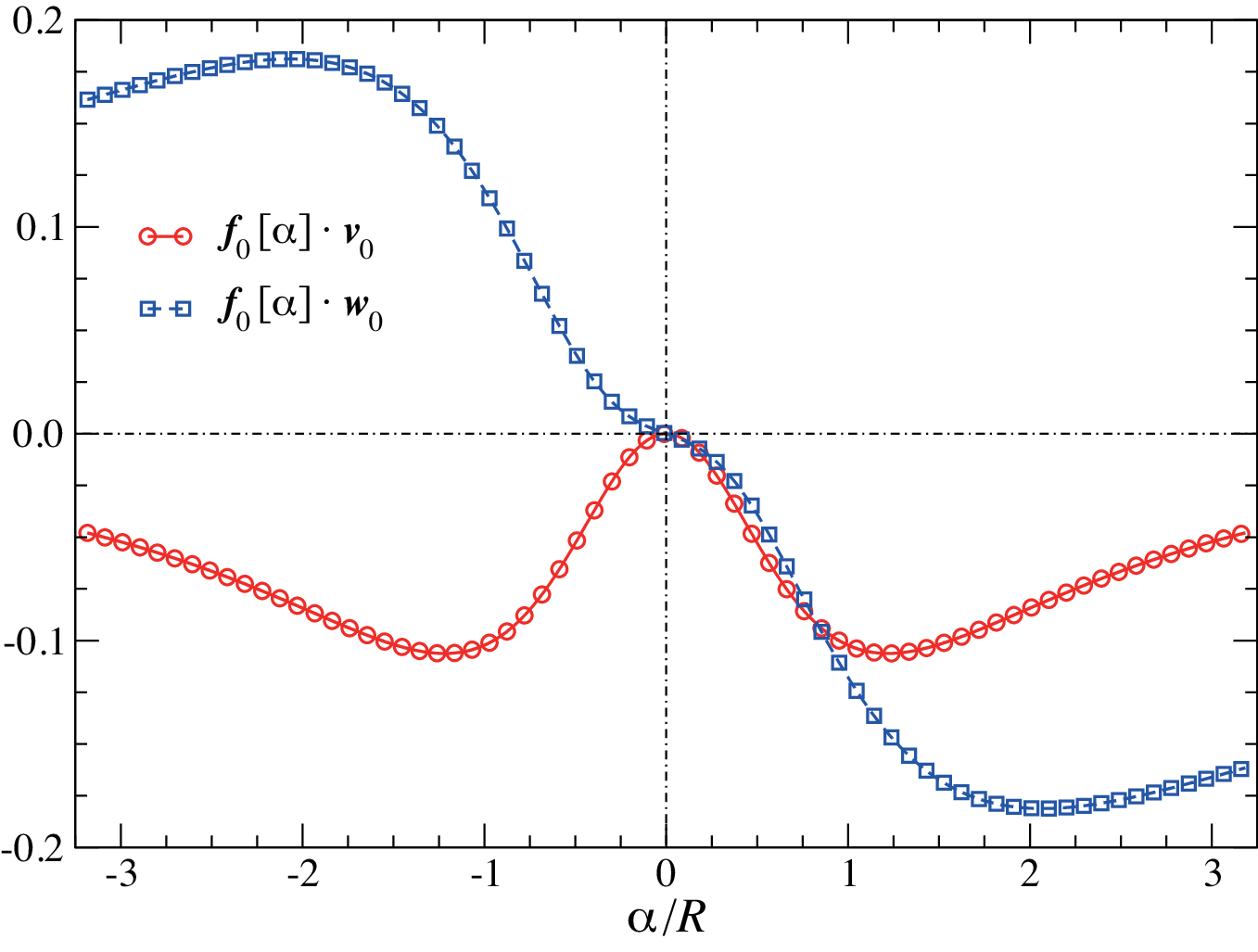}
\caption{The force per length ${\bm f}_0^{~}[ \alpha ]$ in Eq.~(24) acting on the $0$-th wire. 
The circles and squares, respectively, represent the components
${\bm f}_0^{~}[ \alpha ] \cdot {\bm v}_0^{~}$ and
${\bm f}_0^{~}[ \alpha ] \cdot {\bm w}_0^{~}$, measured in units $f^{*}_{~} = \mu N j^2_{~} / 2 \pi R$. }
\end{figure}

\section{Conclusions and Discussions}

We have calculated the magnetic field created by the line currents on the hyperboloid. 
When the tilted angle of the current $\theta$ with respect to the central axis is $\pi / 4$, the electromagnetic force acting on each wire is strongly compensated. 
The fact corresponds to the balance of the Maxwell stress from inside and outside 
of the hyperboloid.

The basic design of the electromagnet we have reported can be considered as 
the starting point for further modifications. For example, we can 
arrange curved wires along the hyperboloid, or we can 
even deform the shape of the
curved surface from the hyperboloid to other negatively curved 
surfaces. A motivation for considering such a modification is to enlarge the space where the magnetic field is strong and quasi-uniform. One of the future theoretical challenges is to find out the wire shape, which perfectly cancels the residual force, which is finite in Fig.~7. 

The force-free nature of the straight current arrangement on the hyperboloid 
could be considered as a fundamental design of high-field magnets. In s realistic experimental setup, the wires have a finite thickness and experience a squashing force from both inside and outside of the hyperboloid. The induced stress is rotationally symmetric with respect to the central axis 
and decays smoothly along the wire, since the strong field region is concentrated
around the origin. The continuous nature would make it easy to support the wires 
mechanically. It should be noted that, like the cylindrical solenoids, it is possible to 
arrange two or more hyperboloid structures in a nested manner.~\cite{Date1,Date2} 

Boundary conditions are another point of consideration. Since it is not possible to
fabricate an infinitely large hyperboloid structure, it is necessary to bend each conducting
wire somewhere. One of the realistic experimental structures, which approximates the hyperboloid under discussion, is the helical coil wound on a torus, as shown in Fig.~8. 
Around the torus hole, the wire is nearly straight and tilted with $\theta$, which
is chosen as $\pi / 4$. Because of the force cancellation around the hole, 
the electromagnetic force imposed on the outer part increases for a certain amount, but the current density in the outer part is small, and also, mechanical support can be performed from a wider area. 
In this direction, to find out an appropriate shape of 
the torus, in particular to adjust the ratio of the hole radius with respect to the whole system size, is the first point to be considered.

\begin{figure}[h]
\centering\includegraphics[width = 7.0 cm]{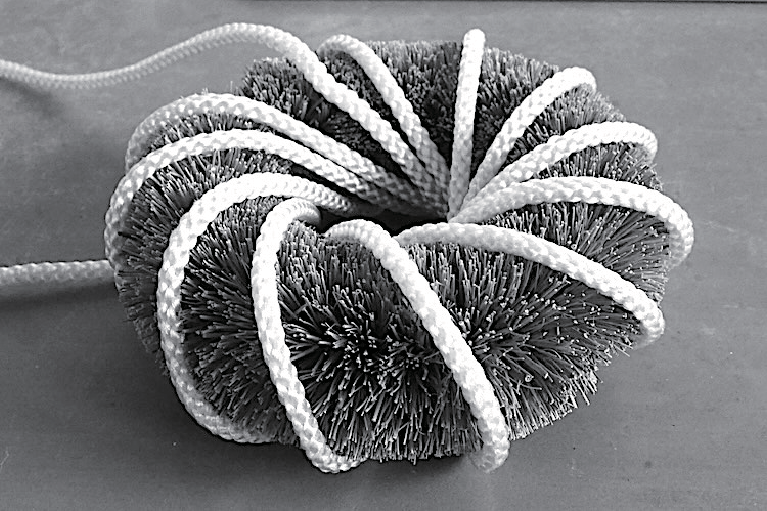}
\caption{The mock-up model of the helical coil wound on the torus.}
\end{figure}

\section*{Acknowledgment}

The authors thank the relaxing landscape of Kobe city, 
which provided an important hint for this work.
This work was supported by the Slovak Research and Development Agency under 
the Contract No. APVV-24-0091, by Vedeck\'{a} Grantov\'{a} Agent\'{u}ra M\v{S}VVaM 
SR and SAV project VEGA No. 2/0152/26.
It was also supported by KAKENHI No. 25K07167.

\let\doi\relax


\begin{thebibliography}{9}
\bibitem{cosh} 
One can introduce the parametrization by means of hyperbolic functions
$\sqrt{x^2_{~} + y^2_{~}} = R \cosh\xi$ and $z \tan\theta = R \sinh\xi$, if necessary.
\end{thebibliography}
\end{document}